\documentclass[preprint2]{aastex}
\usepackage{graphicx}
\usepackage[breaklinks]{hyperref}
\usepackage{breakurl}

\shorttitle{RXJ 0921+4529: a binary quasar or gravitational lens?}
\shortauthors{Popovi\'c et al.}

\begin{document}

 \title{RXJ 0921+4529: a binary  quasar or gravitational lens?}

\author{
L. \v C. Popovi\'c\altaffilmark{1,2}, A.V. Moiseev\altaffilmark{3},
E. Mediavilla\altaffilmark{4},  P. Jovanovi\'c\altaffilmark{1,2}, D.
Ili\'c\altaffilmark{2,5}, J. Kova\v{c}evi\'c\altaffilmark{1,2}, J.
Mu\~noz\altaffilmark{6} }

\altaffiltext{1}{Astronomical Observatory, Volgina 7, 11060
Belgrade, Serbia}

\altaffiltext{2}{Isaac Newton Institute of Chile, Yugoslavia Branch}

\altaffiltext{3}{Special Astrophysical Observatory, Nizhnii Arkhyz,
Karachaevo-Cherkesia, 369167 Russia}

\altaffiltext{4} {Instituto de Astrof\'isica de Canarias, V\'ia
L\'actea S/N, 38200-La Laguna, Tenerife, Spain; Departamento de
Astrof\'isica, Universidad de La Laguna, E-38205 La Laguna,
Tenerife, Spain}

\altaffiltext{5} {Department of Astronomy, Faculty of Mathematics,
Studentski trg 16, Belgrade, Serbia}

\altaffiltext{6} {Departamento de Astronom\'ia y Astrof\'isica,
Universidad de Valencia, 46100-Burjassot, Valencia, Spain}



\begin{abstract}
We report the new spectroscopic observations of the gravitational
lens RXJ 021+4529 with the multi-mode focal reducer SCORPIO of the
SAO RAS 6-m telescope. The new spectral observations were compared
with the previously observed spectra of components A and B of RXJ
0921+4529, i.e. the same components observed in different epochs. We
found a significant difference in the spectrum between the components
that cannot be explained with microlensing and/or spectral
variation. We conclude  that  RXJ 0921+4529 is  a binary quasar
system, where redshifts of quasars A and B are 1.6535$\pm$0.0005 and
1.6625$\pm$0.0015, respectively.
\end{abstract}

\section{Introduction}

Gravitational lenses provide an useful tool for cosmological
investigations, i.e. they can be used to address astrophysical
problems such as the cosmological model, the structure and evolution
of galaxies, and the structure of quasar accretion disks.
Especially, large separated images of a quasar  reveal the
dark-matter content of the lensing galaxies (or galaxy clusters).
However, there are also large separated binary quasar systems that
cannot be interpreted as images of a lensed quasar. Several methods
are developed to confirm or rule out the lens hypothesis for an
observed system \citep[see e.g.][etc]{m98,ko06,mo08}, but, due to
similar quasar spectra, it is sometimes a complex task.

One of the lenses with the large separation between the A and B
images ($6''.97$) is RX J0921+4529 \citep{m01}\footnote{see also at
\url{http://www.cfa.harvard.edu/castles/}}. \citet{m01} reported
multiwavelength observation of RX J0921+4529 finding that the system
contains two images of a quasar at $z_s=$ 1.66. They also observed a
spiral galaxy between the quasar images, which is probably a member
of an X-ray cluster at  $z_l=$0.32. It was interesting that an
extended source was detected near the fainter quasar image B (denoted as B$'$), but
not in the  image A \citep{m01}. If this extended source around
image B corresponds to the quasar host then the system would be a
binary  quasar rather than a gravitational lens. Moreover,
\citet{p06} found that RX J0921+4529 is a binary quasar rather than
a gravitational lens. They  analyzed the host galaxies of lensed
quasars and in RX J0921+4528 they did not find any effects of an
Einstein ring structure to the host galaxies. Also, when they
modeled the system as a lens, they found that the host galaxy has a
huge inferred mass deficit (around 7-8 times more than expected),
while when they treated the system as a binary quasar they found a
mass deficit that is typical for the other host galaxies at that
redshift.

In principle, there may be several reasons for difference between
the spectra of the images of a lensed quasar \citep{pc05}: (i) in
the continuum/line it may be caused by the gravitational
microlensing \citep[see e.g.][etc.]{a02,sl07,m09}; (ii) by the
intrinsic variations, (iii) extinction can cause difference in the
line profiles and in the continuum shapes \citep[see
e.g.][]{m04,pc05}; and finally (iv) there is a small probability,
but nevertheless it is also possible, that an image (in this case B)
is projected very close to another object with emission lines, and
moreover, \citet{m01} reported about an extensive source around the
image B of RX J0921+4529.

The spectra of active galactic nuclei (AGN) can show a very  high
variability, not only in the continuum but also in line shapes
\citep[as e.g. in the case of NGC 4151, see][]{sh08,sh09}. In the
case of intrinsic variability, one can expect that the spectra from
two epochs (as well as for both images) are  similar \citep[see
e.g.][]{s97}. For a lens with a time delay $\sim$100 days, as it was
estimated for RX J0921+4529 by \cite{m01}, the observed velocity
difference could be created by quasar variability coupled with a
long time delay. Moreover, the extended object near
the B quasar may be a faint galaxy rather than the host galaxy of
quasar B, therefore the system can be a lens with large separation
of images.

In order to clarify the nature of the RX J0921+4529 we performed the
new spectroscopic observations of this system with the 6-m telescope
of SAO, using the long-slit spectroscopy. The observed spectra were
compared with those published in \citet{m01}. In this letter we
report our observations (in \S 2), we discuss obtained results (in
\S 3) and outline our conclusions (in \S 4).

\section{Observations and data reduction}

Long-slit spectral observations were performed on October 29/30,
2008 (hereafter spectra from epoch 2) with the multi-mode focal
reducer SCORPIO \citep{AfanasievMoiseev05} installed at the prime
focus of the BTA~6-m telescope at the Special Astrophysical
Observatory of the Russian Academy of Sciences.  The seeing was
$1\farcs2-1\farcs4$.  A $1''$ wide slit was placed along A and B
components of RXJ~0921+4529 at the position angle $PA=115^\circ$.
The spectral range was 3650--7540\,\AA\ with a spectral resolution
8-10\,\AA\ FWHM. With a CCD EEV 42-40 $2048\times2048$ pixels
detector, the reciprocal dispersion was $1.9$\,\AA\ per pixel. The
total exposure time was 9600\,s, divided into eight 20-minute
exposures. The target was moved along the slit between exposures to
ease background subtraction and CCD fringes removal in the data
processing.  The bias subtraction, geometrical corrections, flat
fielding, sky subtraction, and calibration to flux units
($F_{\lambda}$) was performed by means of IDL-based software shortly
described in \citet{AfanasievMoiseev05}.

To compare the spectra of images between two different epochs we
used the already published spectra of A and B obtained with the MMT
and Blue Chanel spectrograph \citep[hereafter spectra from epoch 1,
for more details see Fig. 2 and corresponding text in][]{m01}.

\begin{figure}[ht!]
\begin{center}
\includegraphics[width=\columnwidth]{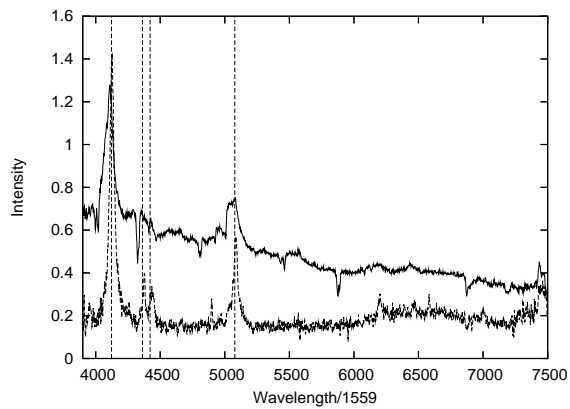}
\caption{The observed spectra of components A (solid
line) and B (dashed line), the intensity of component B was
multiplied five times for the comparison of the spectra. The
intensity is given in $10^{-16}\rm erg \ cm^{-2}s^{-1}$. Position of
CIV$\lambda$1550, HeII$\lambda$1640, OIII$\lambda\lambda$1662 and
CIII]$\lambda$1909 redshifted at 1.66 are denoted with vertical
dashed lines.}
\label{f01}
\end{center}
\vspace*{-0.5cm}
\end{figure}

\begin{figure*}[ht!]
\begin{center}
\includegraphics[width=\columnwidth]{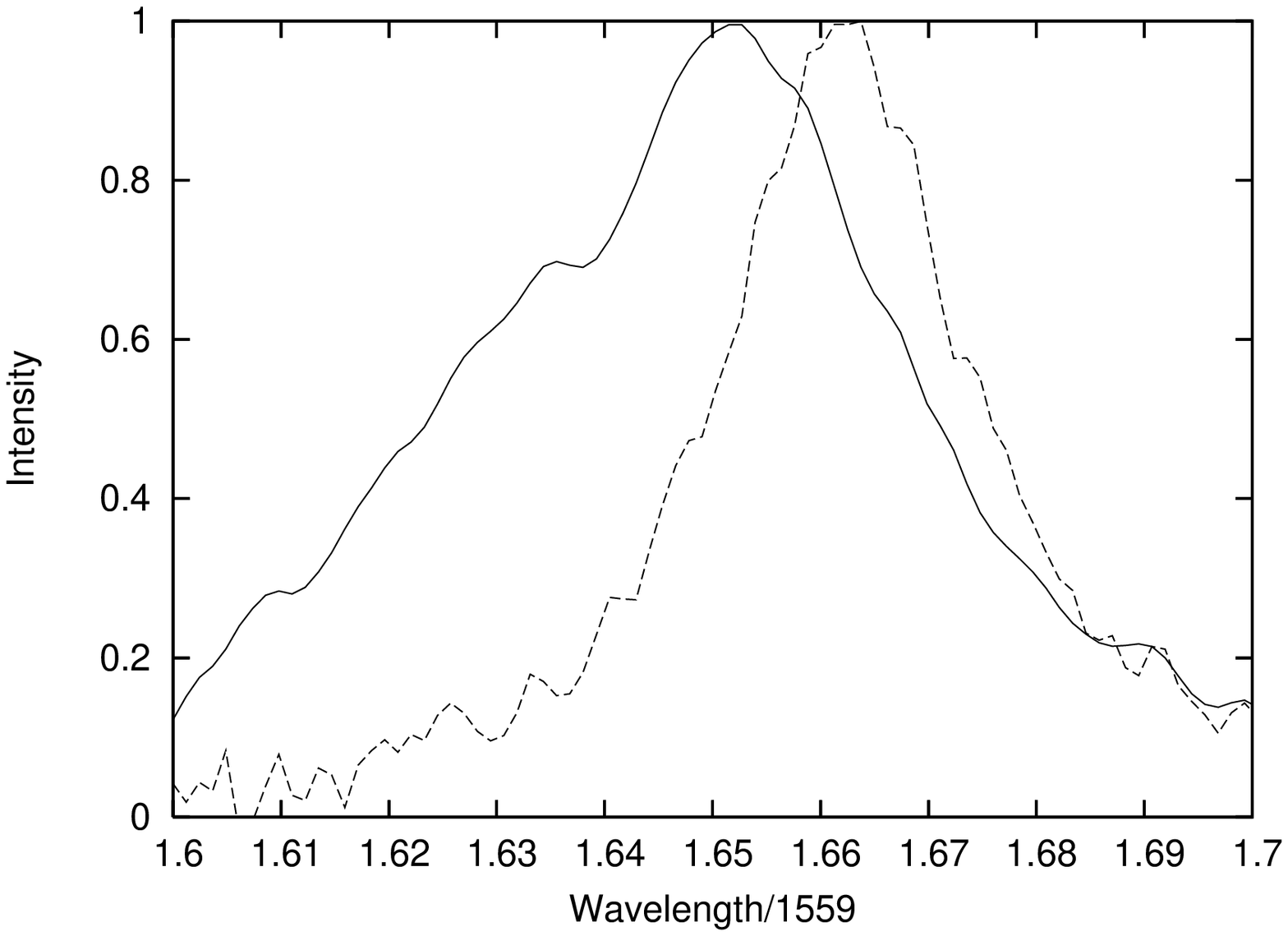}
\includegraphics[width=\columnwidth]{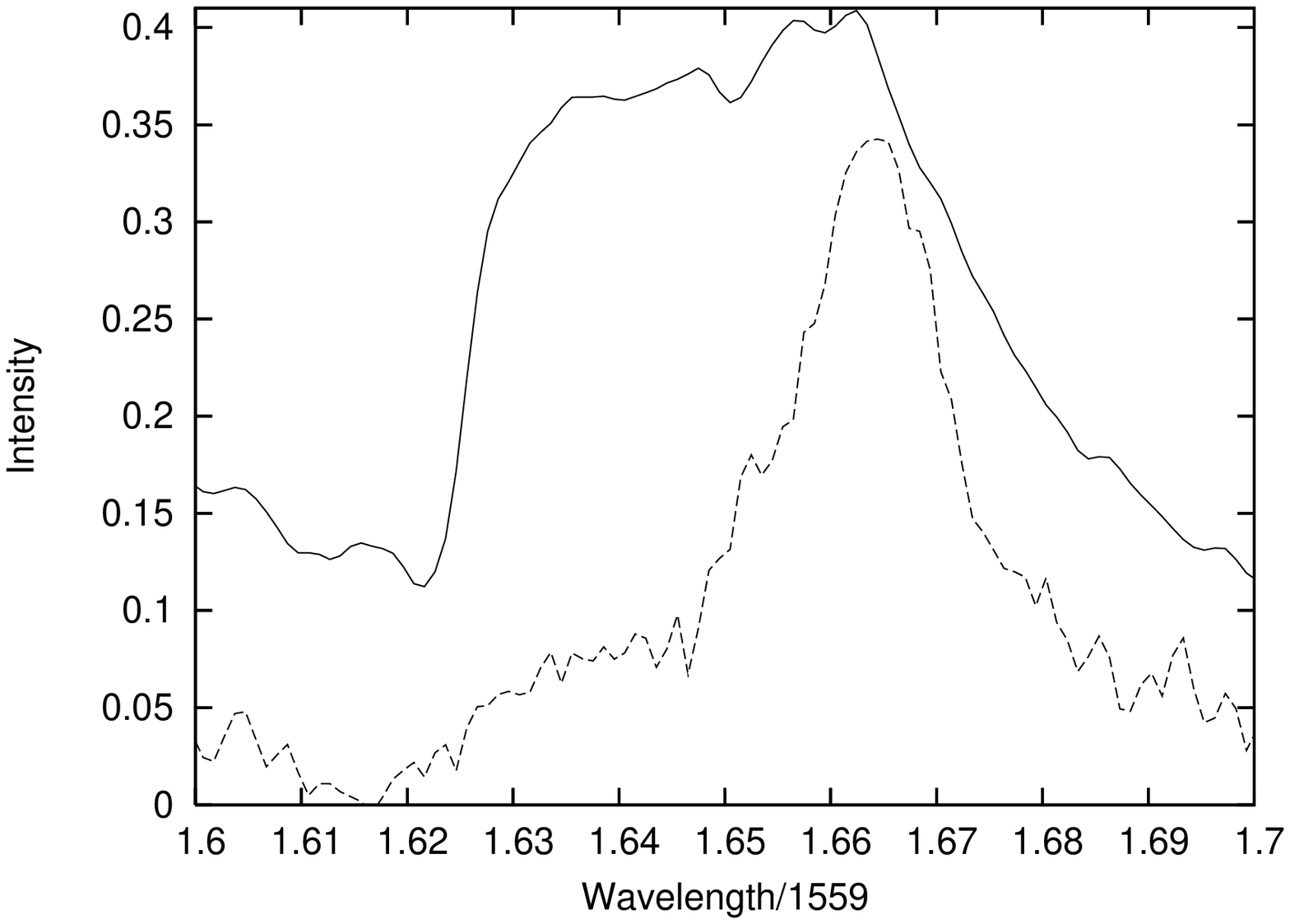} \\
\includegraphics[width=\columnwidth]{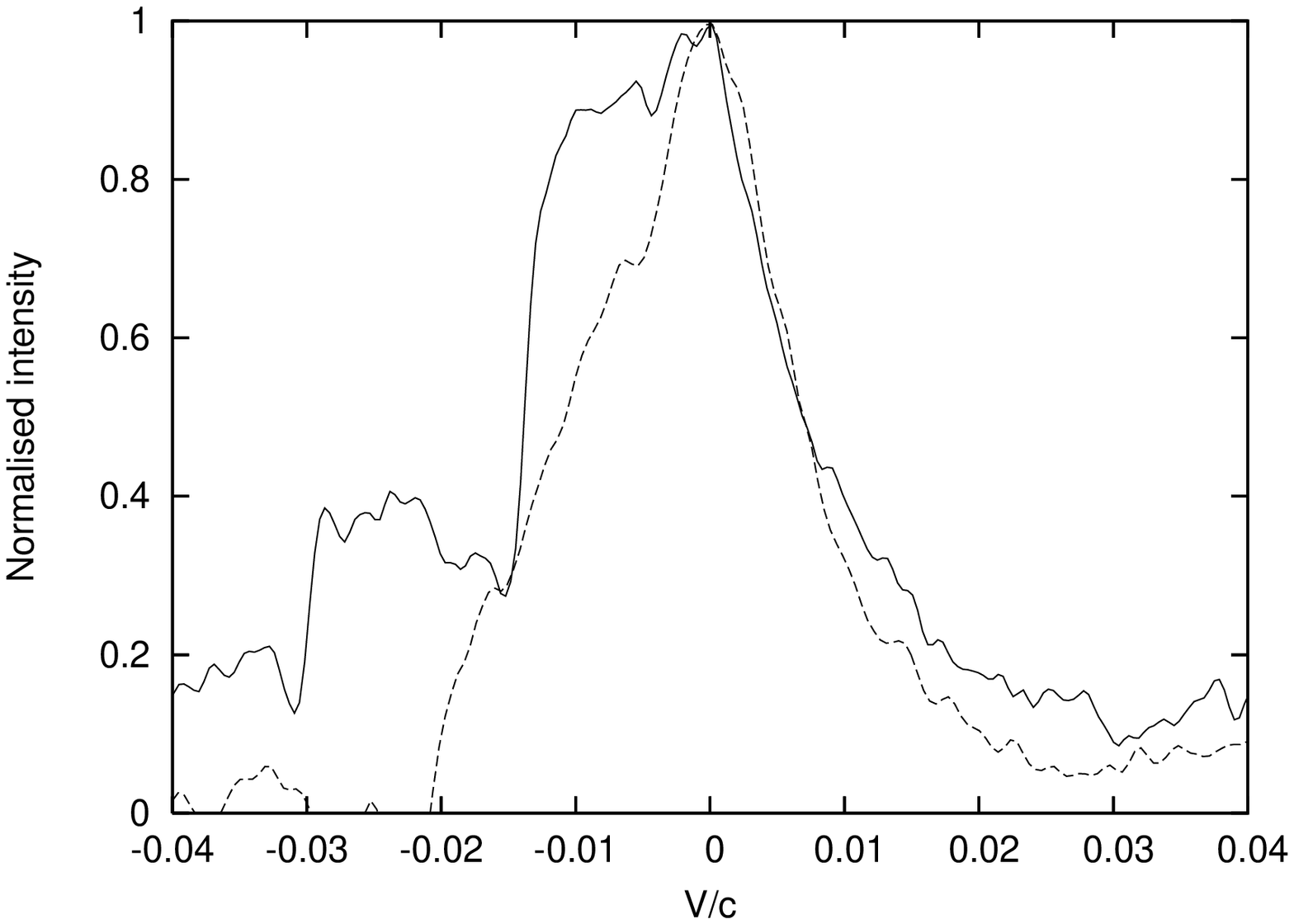}
\includegraphics[width=\columnwidth]{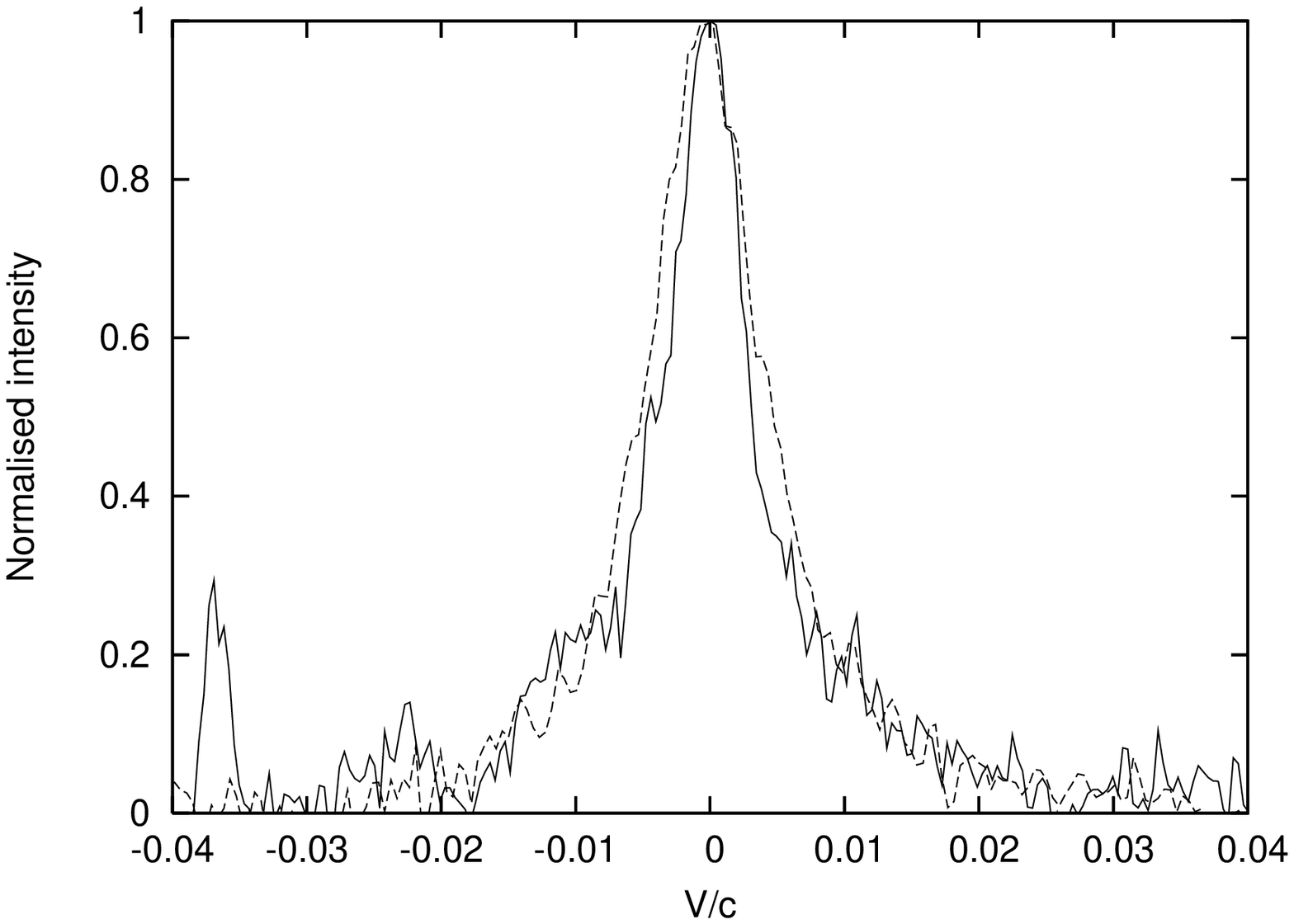}
\caption{Top left: the comparison of the C IV line of the component
A (solid line) and component B (dashed line) in the scale of the
redshift; Top right: the same as in left Fig, but for CIII] line;
Bottom left: the comparison of the C IV (dashed line) and CIII]
(solid line) line profiles of the component A; Bottom right: the
same as in left Fig, but for component B.}
\label{f02}
\end{center}
\vspace*{-0.5cm}
\end{figure*}

\section{Results and discussion}

\subsection{Analysis of long-slit spectra}

First inspection of the RX J0921+4529 A and B component spectra show
a big difference between the lines as well as the continuum (see
Fig. \ref{f01}). The lines of component B are narrower than those of
the component A. Additionally in the spectrum of the component B
there are prominent HeII$\lambda$1640 and
OIII$\lambda\lambda$1661,1663 emission lines which are not present
(or they are too weak) in component A. On the other hand, it seems
that Si III]$\lambda$1892 line in the blue wing of
CIII]$\lambda$1909 is more intensive in the component A than in B of
the system.

Next, we measured the line parameters of the most intensive C IV
line in the spectra of  both components, and also, we measured the
redshift from the line peak. We found the redshifts 1.654 for
component A and 1.664 for component B. Also the line widths of C IV
line (Full Width at Half Maximum Intensity - FWHM) are quite
different, i.e. FWHM of C IV in component A is 5300 kms$^{-1}$, and
in component B is 3000  kms$^{-1}$. The lines in component A show a
blue asymmetry, while in component B a red one.

Additionally, we compared line profiles of C IV lines from component
A and B (see Fig. \ref{f02} top) and found that line profiles are
different between components; also, the line profiles of C IV and
CIII] are the same in component B (see Fig. \ref{f02} bottom right), while
they  are quite different in component A (see Fig. \ref{f02} bottom left),
the difference in the blue wing of C IV and CIII] in the component A
may be caused by the contribution of the Si III]$\lambda$1892 line.

\begin{figure*}[ht!]
\begin{center}
\includegraphics[width=\columnwidth]{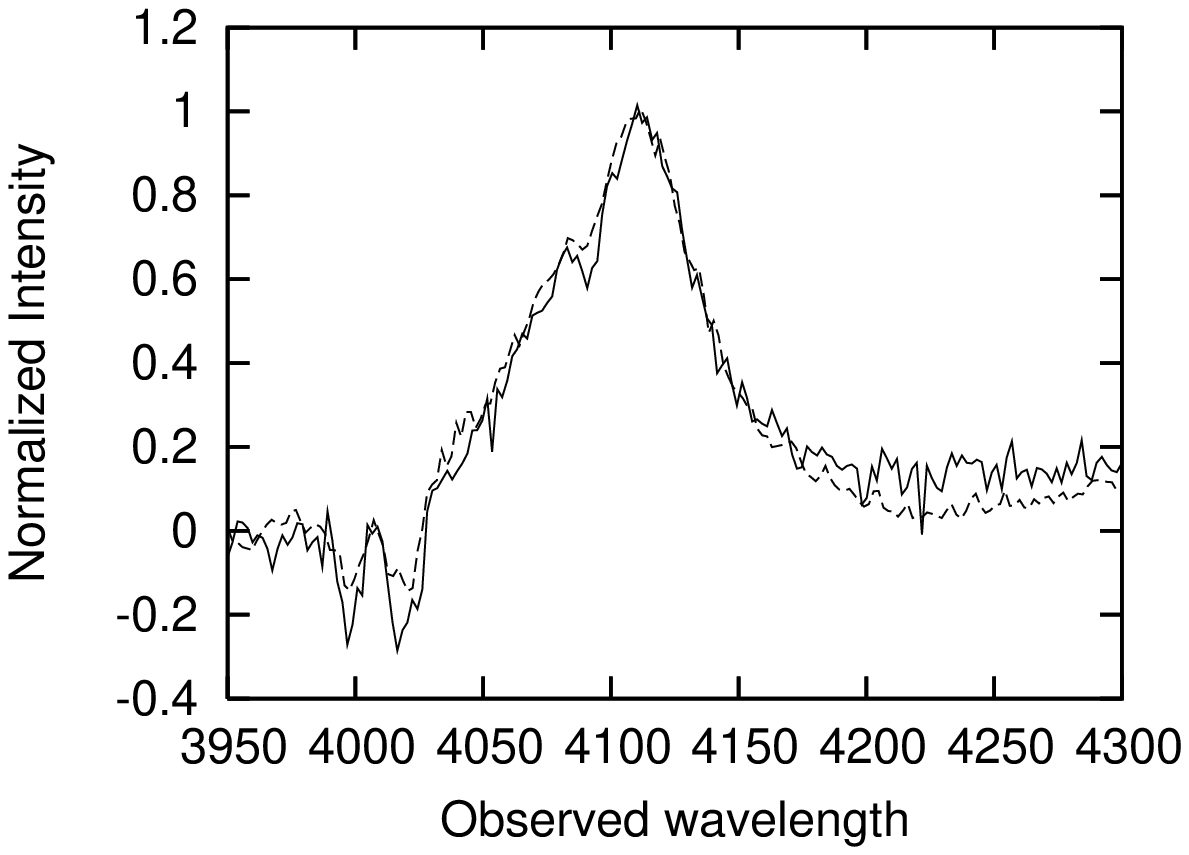}
\includegraphics[width=\columnwidth]{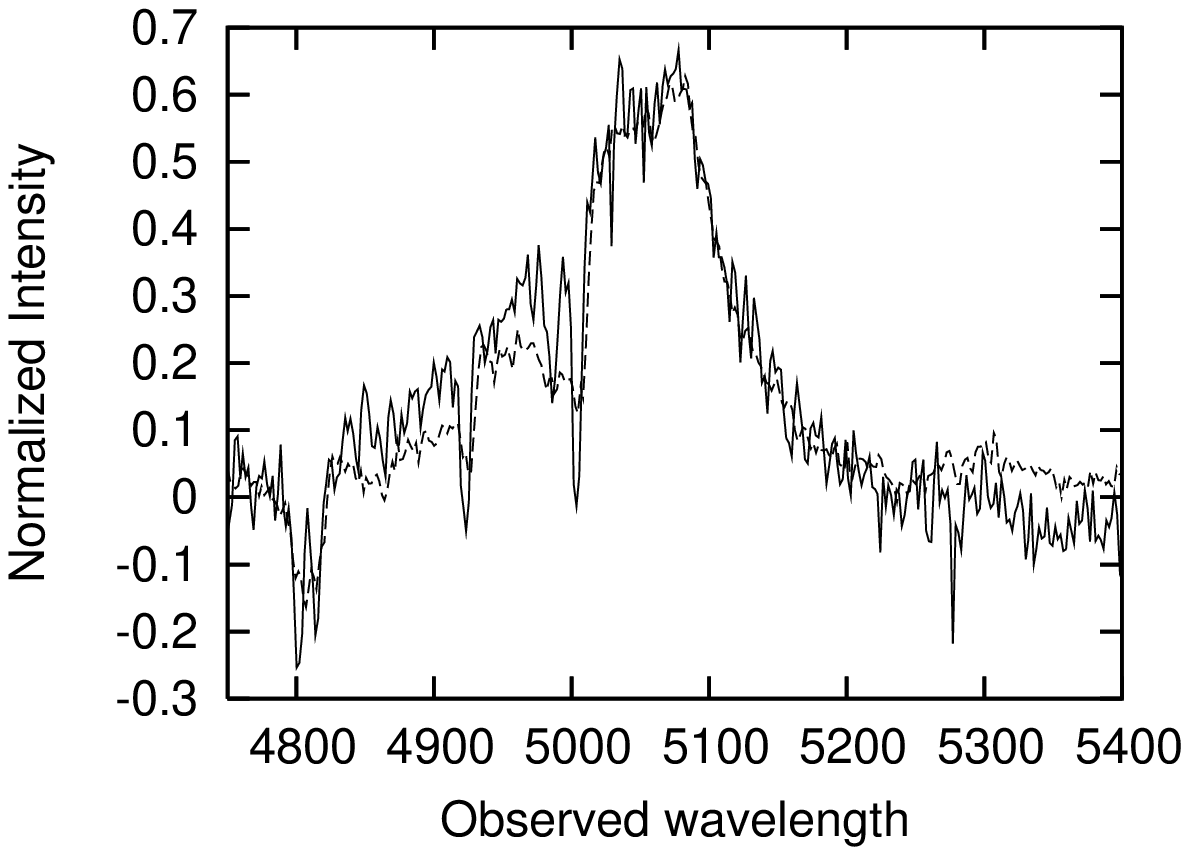} \\
\includegraphics[width=\columnwidth]{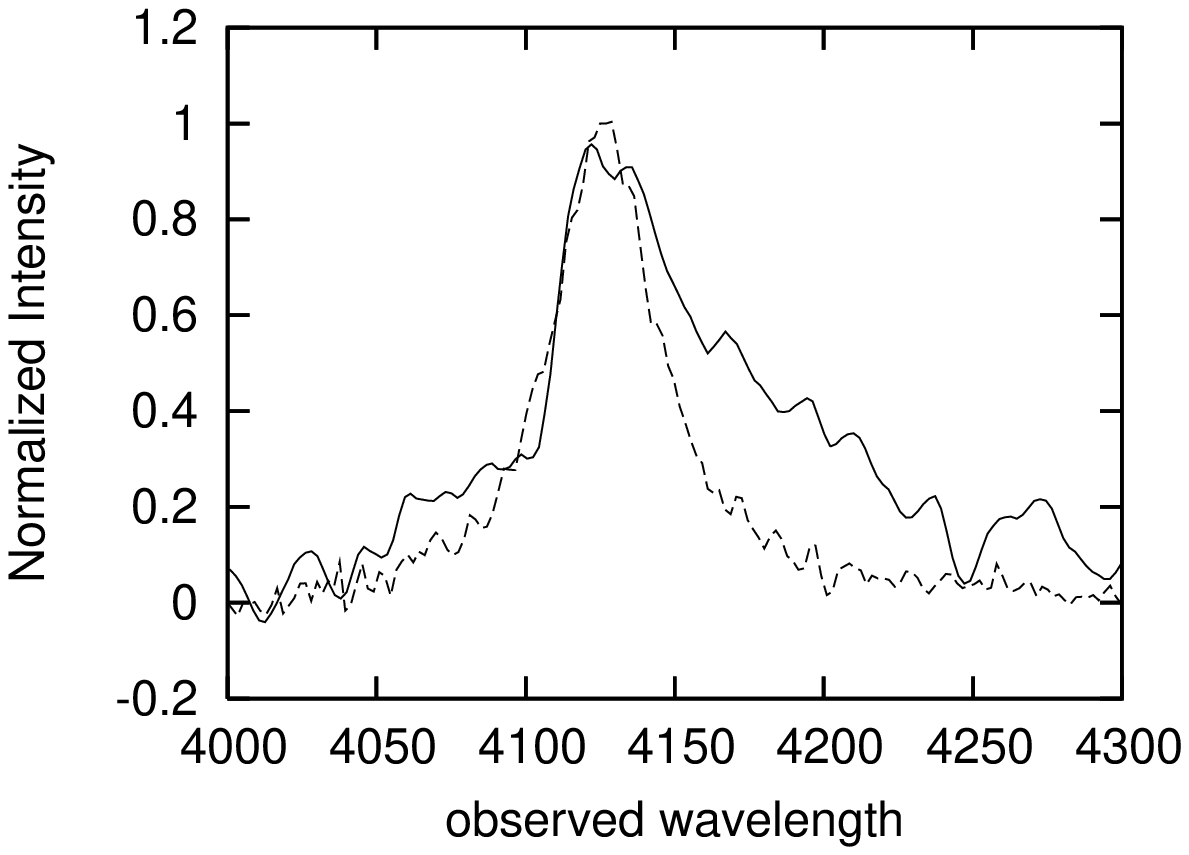}
\includegraphics[width=\columnwidth]{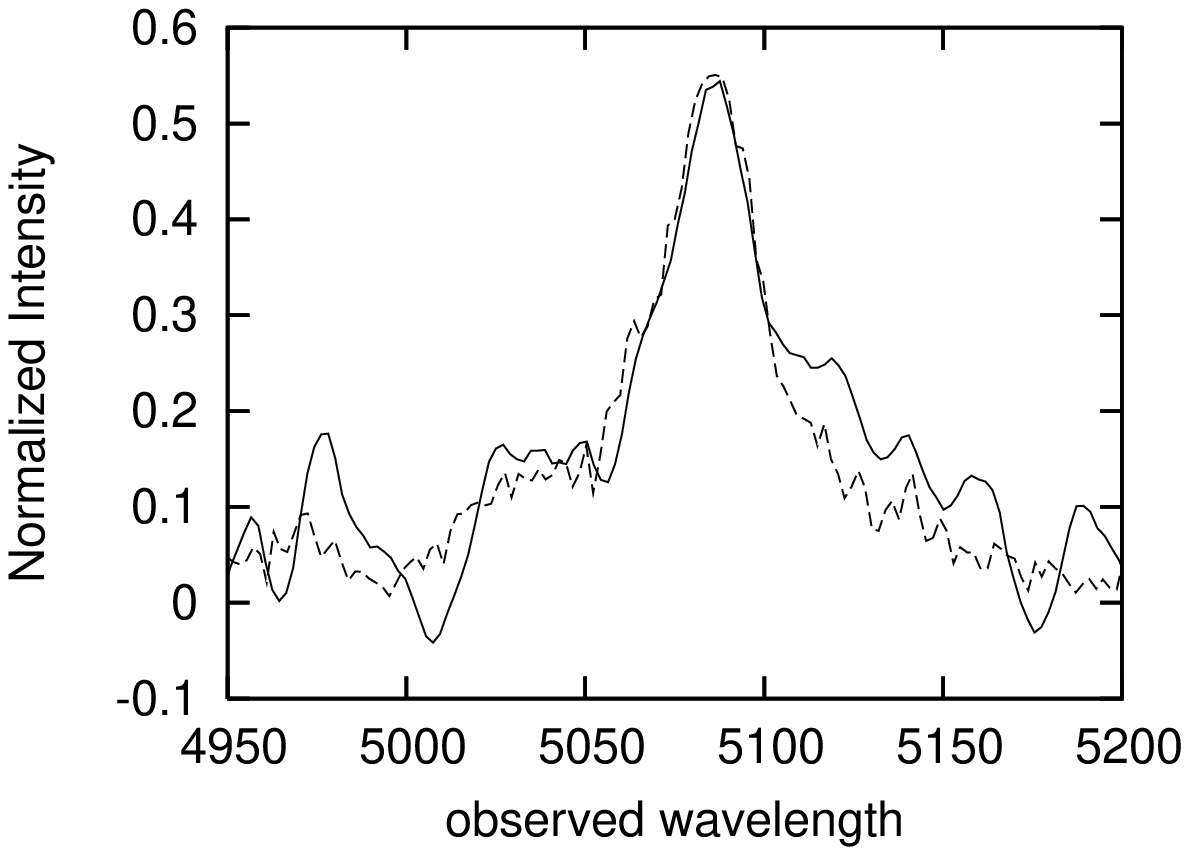}
\caption{Comparison between the C IV and CIII] line shapes observed
in two epochs, corresponding to the component A (top) and to the
component B (bottom). Solid lines represent observations from epoch
1 and dashed from epoch 2.}
\label{f03}
\end{center}
\end{figure*}

\subsection{Comparison of long-slit and MMT spectra}

First we subtracted continuum in spectra of both components in both
epochs, then normalized the spectra on the maximum intensity of the
C IV line. By comparing the spectra between the same images obtained
from two different epochs (see Fig. \ref{f03}) we found that the
line shapes in component A from the two epochs are similar (there
are differences in intensity that may be caused by variability), but
in the case of the component B the lines observed in epoch 1 have a
stronger red asymmetry (stronger red wing) than ones observed in
2008 (especially in the CIV line). Also, we found different
redshifts in components A and B which are 1.653 and 1.661,
respectively. That is close to measured redshifts of components A
and B in epoch 2. From two epochs we found that the redshifts of the
components A and B are significantly different (averaged
1.6535$\pm$0.0005 for component A and 1.6625$\pm$0.0015 for
component B), as well as other parameters in both epochs, implying
that source of radiation is not identical. The difference between
components A and B cannot be explain by intrinsic variability, and
therefore we can conclude that the system is rather binary quasar
than gravitational lens. Taking the average redshift of A and B
quasars ($z=1.658$) and assuming flat cosmological model with
$\Omega_\mathrm{m}$­ = 0.27, $\Omega_\mathrm{\Lambda} = 0.73$ and
$H_0 = 71$ km s$^{-1}$ Mpc$^{-1}$, the corresponding angular
diameter distance of RX J0921+4529 is $1765$ Mpc and the transverse
separation (i.e. projected linear distance) between A and B
components is 59.6 kpc.

\begin{figure*}[ht!]
\begin{center}
\includegraphics[width=\textwidth]{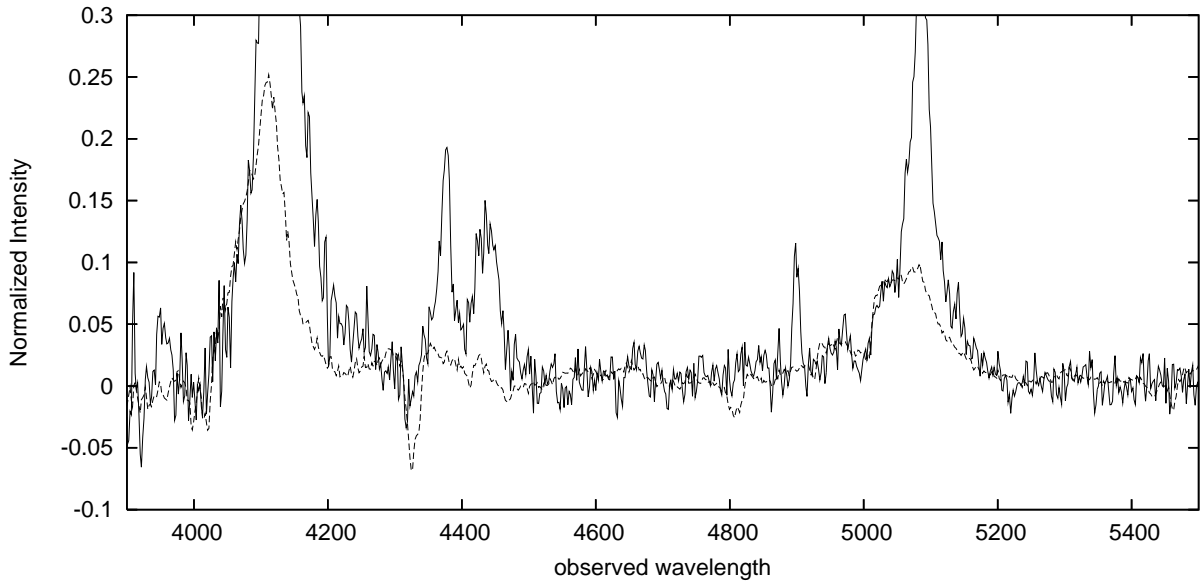}
\caption{ Comparison of the line spectrum of the component A (dashed
line) and B (solid line) observed in 2008, where the line
intensities (normalized on maximum of C IV) of component A are
divided by 4.}
\label{f04}
\end{center}
\end{figure*}

\begin{figure}[ht!]
\begin{center}
\includegraphics[width=\columnwidth]{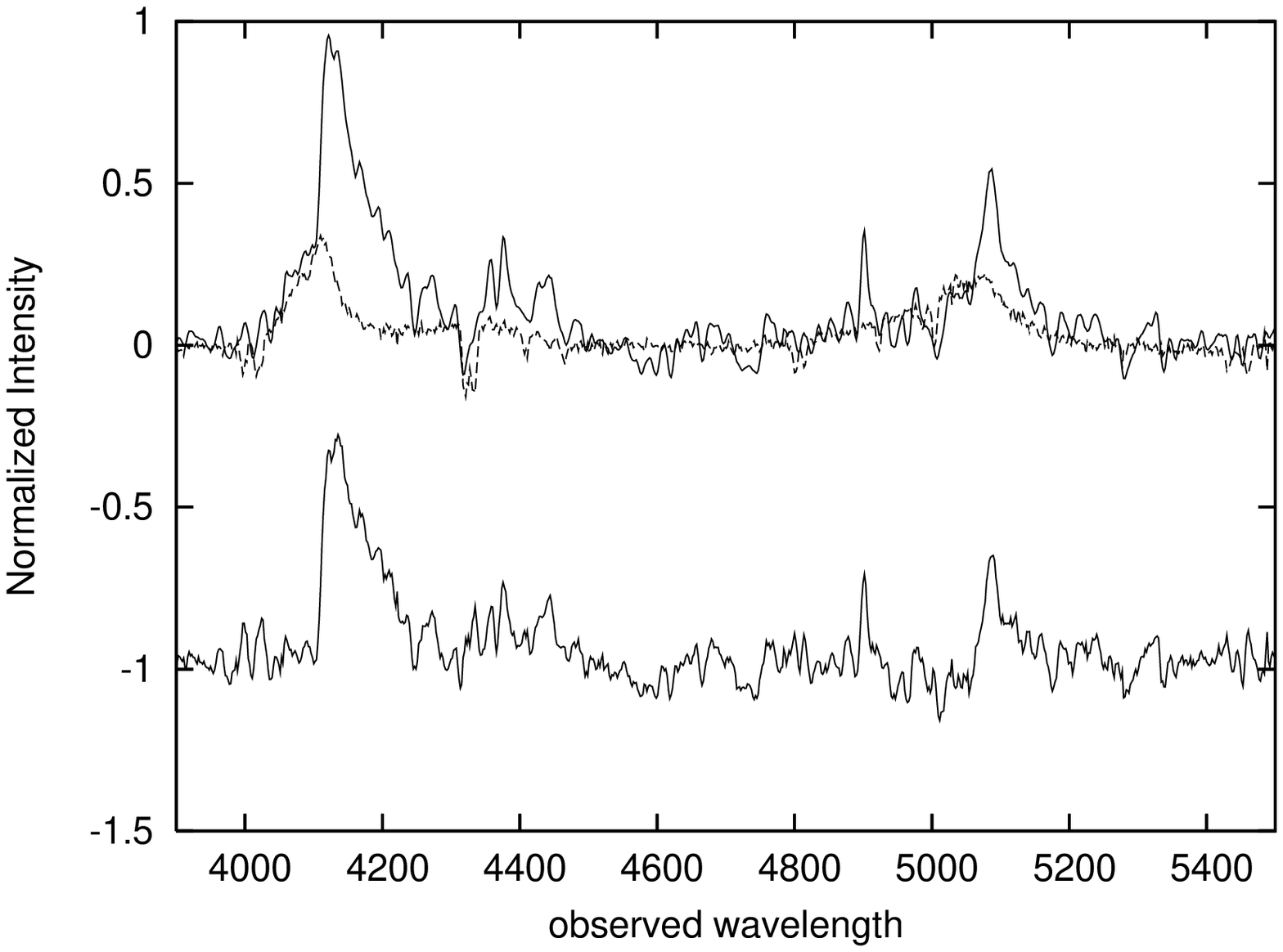} \\
\includegraphics[width=\columnwidth]{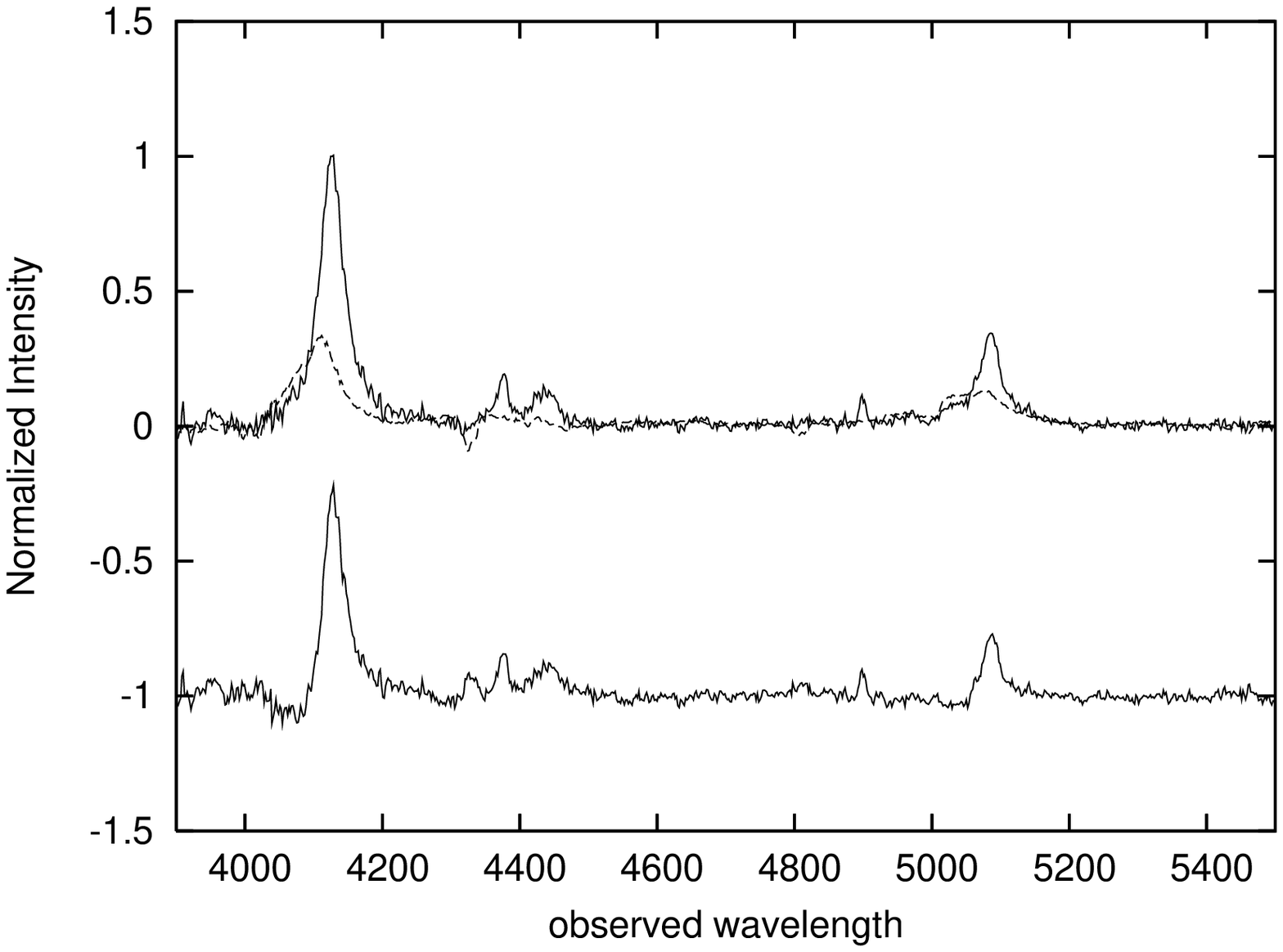} \\
\includegraphics[width=\columnwidth]{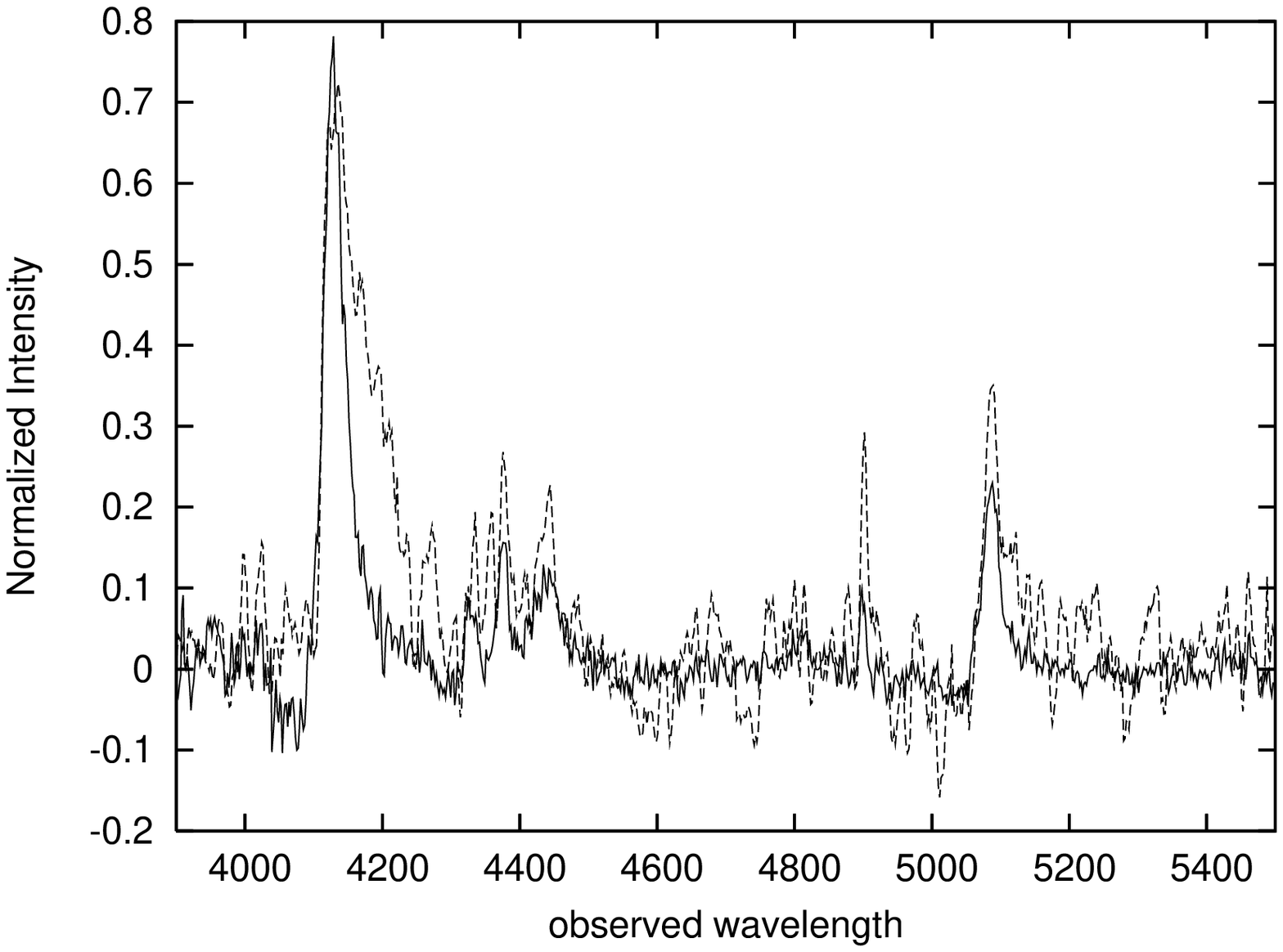}
\caption{The component B line spectrum when we assume that four times
weaker emission from the component A is present in the spectra of the
component B (two first panels, epochs 1 and 2 respectively). The third
panel presents comparison between component B spectra in epochs 1 (dashed line)
and 2 (solid line) when assumed that contribution of component A is subtracted.}
\label{f05}
\end{center}
\end{figure}

Note here, that the spectrum of component B may be composed from the
light of two sources, since \citep{m01} reported about a extended
source near the component B (denoted as B', see also observations at
\burl{http://www.cfa.harvard.edu/castles/Individual/RXJ0921.html}).
Then, one hypothesis may be that  the spectrum of component B is
composed from the light of one component of lensed QSO (the same as
component A) and from the emission of an additional source. To check
it, we tried to fit the spectra of component A into spectra of
component B, and we found that a weak spectral component of A may be
present in the component B (see Fig. \ref{f04}). We multiplied the
intensity (normalized to the C IV line maximal intensity) of
component A with 0.25 and found that it fitted well the blue wings
of lines observed in the component B. After subtraction of the A
spectral component in both epoch (see Fig. \ref{f05}) we obtained
that the red asymmetry observed in epoch 1 remained. It means that
there is line shape variability present only  in the component B. Of
course, the spectra of quasars are very similar, and this should be
taken with caution. The future observations are needed to
distinguish if RXJ 0921+4529 is an ordinary binary quasar or an
unique object.

\section{Conclusion}

In this paper we report briefly the spectroscopic observations of
the system RXJ 0921+4529. From our spectral observations of the
system, we can conclude that the spectral properties (line
parameters) of RXJ 0921+4529 A and B components are quite different
indicating that it is a binary quasar. The intrinsic variability
coupled with a long time delay cannot explain such difference in
spectra of the components A and B. Also, the future precise spectral
observations of components B and B' would give more information
about the structure of the system, that may be more complex than an
ordinary binary quasar.

\section*{Acknowledgments}

This work is  based on the observational data obtained with the 6-m
telescope of the Special Astrophysical Observatory of the Russian
Academy of Sciences funded by the Ministry of Science of the Russian
Federation (registration number 01-43). This work was supported by
the Ministry of Science of Serbia through the project "Astrophysical
Spectroscopy of Extragalactic Objects". Also, the work has been
financed by the Russian Foundation for Basic Research (project no.~
06--02--16825). A.V.M. is grateful to "Dynasty" foundation. The
authors would like to thank the anonymous referee for very useful
comments.

\clearpage

\end{document}